\documentclass[aps,prl,twocolumn,showpacs,amsmath,amssymb, nofootinbib, noeprint]{revtex4-1}
\usepackage{epsfig}
\usepackage{graphicx}
\usepackage{dcolumn}
\usepackage{bm}
\usepackage{ltablex,booktabs}
\usepackage{overpic}
\usepackage{subfigure}
\usepackage{float}
\usepackage{color}
\usepackage{amsmath}
\usepackage{mathcomp}
\usepackage{mathrsfs}
\usepackage{multirow}
\usepackage{rotating}
\usepackage{amssymb}
\usepackage{gensymb}
\usepackage{amsmath}
\usepackage{tabularx}
\usepackage{lineno}
\usepackage{tikz}
\usepackage[compat=1.1.0]{tikz-feynman}
\tikzfeynmanset{warn luatex=false}
\usepackage[bookmarksnumbered, pdfstartview=FitH,colorlinks,urlcolor=blue, citecolor=blue,linkcolor=blue] {hyperref}

\newcommand{\BESIIIorcid}[1]{\href{https://orcid.org/#1}{\hspace*{0.1em}\raisebox{-0.45ex}{\includegraphics[width=1em]{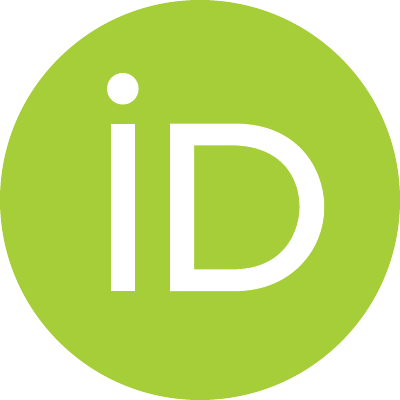}}}}

\begin{document}
\normalsize
\parskip=5pt plus 1pt minus 1pt


\title{
First Observation of \boldmath{$D^+ \to a_0(980)\rho$ and $D^+ \to a_0(980)^+ f_0(500)$} \\
in \boldmath{$D^+ \to \pi^+\pi^+\pi^-\eta$ and $D^+ \to \pi^+\pi^0\pi^0\eta$} Decays
}

\author{
\begin{small}
  \begin{center}
M.~Ablikim$^{1}$\BESIIIorcid{0000-0002-3935-619X},
M.~N.~Achasov$^{4,c}$\BESIIIorcid{0000-0002-9400-8622},
P.~Adlarson$^{81}$\BESIIIorcid{0000-0001-6280-3851},
X.~C.~Ai$^{87}$\BESIIIorcid{0000-0003-3856-2415},
C.~S.~Akondi$^{31A,31B}$\BESIIIorcid{0000-0001-6303-5217},
R.~Aliberti$^{39}$\BESIIIorcid{0000-0003-3500-4012},
A.~Amoroso$^{80A,80C}$\BESIIIorcid{0000-0002-3095-8610},
Q.~An$^{77,64,\dagger}$,
Y.~H.~An$^{87}$\BESIIIorcid{0009-0008-3419-0849},
Y.~Bai$^{62}$\BESIIIorcid{0000-0001-6593-5665}
% ...
\\
\vspace{0.2cm}
(BESIII Collaboration)\\
\vspace{0.2cm} {\it
$^{1}$ Institute of High Energy Physics, Beijing 100049, People's Republic of China\\
% ...
$^{4}$ Budker Institute of Nuclear Physics SB RAS (BINP), Novosibirsk 630090, Russia\\
% ...
$^{31}$ INFN Sezione di Ferrara, (A)INFN Sezione di Ferrara, I-44122, Ferrara, Italy; (B)University of Ferrara, I-44122, Ferrara, Italy\\
% ...
$^{39}$ Johannes Gutenberg University of Mainz, Johann-Joachim-Becher-Weg 45, D-55099 Mainz, Germany\\
% ...
$^{62}$ Southeast University, Nanjing 211100, People's Republic of China\\
% ...
$^{64}$ State Key Laboratory of Particle Detection and Electronics, Beijing 100049, Hefei 230026, People's Republic of China\\
% ...
$^{77}$ University of Science and Technology of China, Hefei 230026, People's Republic of China\\
% ...
$^{80}$ University of Turin and INFN, (A)University of Turin, I-10125, Turin, Italy; (B)University of Eastern Piedmont, I-15121, Alessandria, Italy; (C)INFN, I-10125, Turin, Italy\\
$^{81}$ Uppsala University, Box 516, SE-75120 Uppsala, Sweden\\
% ...
$^{87}$ Zhengzhou University, Zhengzhou 450001, People's Republic of China\\

\vspace{0.2cm}
$^{\dagger}$ Deceased\\
% ...
$^{c}$ Also at the Novosibirsk State University, Novosibirsk, 630090, Russia\\
% ...

}

\end{center}
\end{small}
}

  
\begin{abstract}
We perform the first amplitude analysis of the singly Cabibbo-suppressed decays $D^+ \to \pi^+ \pi^{+(0)} \pi^{-(0)} \eta$,
using  $e^+e^-$ collision data collected with the BESIII detector at the center-of-mass energy of 3.773\,GeV, corresponding to an integrated luminosity of 20.3 $\rm{fb}^{-1}$.
The absolute branching fractions of the $D^+ \to \pi^+ \pi^+ \pi^- \eta$ and $D^+ \to \pi^+ \pi^0 \pi^0 \eta$ decays are measured to be $(3.20\pm0.06_{\text{stat.}}\pm0.03_{\text{syst.}})\times 10^{-3}$ and $(2.43 \pm 0.11_{\text{stat.}} \pm 0.04_{\text{syst.}}) \times 10^{-3}$, respectively.
The decay process $D^{+}\to a_0(980)^{+}f_0(500)$ is observed for the first time with an unexpectedly large branching fraction.
Moreover, we observe the decays $D^+ \to a_0(980)^{+(0)} \rho(770)^{0(+)}$ and measure the ratio $r_{+/0} \equiv \frac{\mathcal{B}(D^+ \to a_0(980)^+ \rho(770)^0)}{\mathcal{B}(D^+ \to a_0(980)^0 \rho(770)^+)}$ for the first time to be $0.55\pm0.08_{\text{stat.}}\pm0.05_{\text{syst.}}$.
These results offer a novel insight into our comprehension of the nature of the $a_0(980)$ and $f_0(500)$ states.

\end{abstract}

\maketitle

Although the existence of light scalar mesons, particularly the $f_{0}(500)$, $f_{0}(980)$, and $a_{0}(980)$, has been firmly established experimentally, their classification within the constituent quark model~\cite{PDG} remains a subject of intense debate. These states present significant theoretical challenges when attempting to describe them as conventional quark-antiquark configurations, such as the heavy mass of $a_0(980)$~\cite{PDG}, the observation of the Okubo-Zweig-Iizuka suppressed decay $\phi \to \gamma a_{0}(980)^0$~\cite{Braghin:2022uih}, and several unexpected large values of branching fraction~(BF) for decays involving scalar mesons~\cite{BESIII:2018sjg,BESIII:2019jjr,BESIII:2021aza,BESIII:2024tpv}. 
These intriguing phenomena motivate the development of alternative hypotheses on their internal structure, including compact tetraquark states~\cite{Hsiao:2023qtk, Hsiao:2019ait,Yu:2021euw,Jaffe:1976ig,Brito:2004tv,Klempt:2007cp,Alexandrou:2017itd,Humanic:2022hpq}, two-meson molecule bound states~\cite{Weinstein:1982gc,Dai:2014lza,Sekihara:2014qxa,Duan:2020vye, Ikeno:2021kzf, Ke:2023qzc}, and mixed states~\cite{Braghin:2022uih}. The production of these exotic structures typically relates to final state interactions~(FSIs), whose mechanisms are predominantly governed by non-perturbative QCD. 

\begin{figure}[htbp]
  \vspace{-2.5em}
  \centering
          \mbox{
            \put(-140, 0){
            \includegraphics[width=0.29\textwidth]{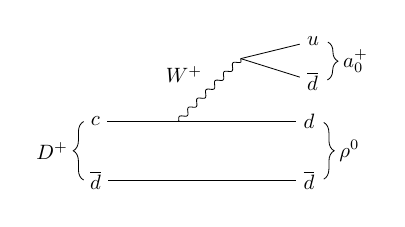}
            }
            \put(-15, 0){
            \includegraphics[width=0.29\textwidth]{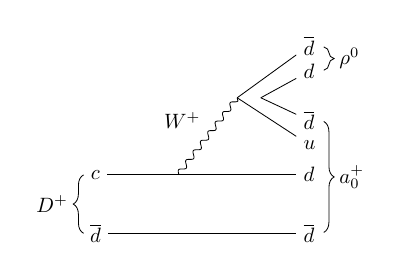}
            }
            \put(-80, 0) { $(a)$}
            \put(70, 0)  {$(b)$}
        }
  \caption{Topological diagrams for the decay $D^+ \to a_0(980)^+ \rho(770)^0$ showing the external $W$-emission process with $a_0(980)^+$ (a) as a two-quark state, and (b) as a tetraquark state. (b) shows one of the several possible diagrams. Other diagrams contributing to related decays are discussed in Ref.~\cite{Cheng:2024zul}.
  }
  \label{fig:feynman_diagram}
\end{figure}

Complementarily, it is known that hadronic decays of charmed mesons receive significant contribution from FSIs, providing an exceptional laboratory for investigating the nature of light scalar mesons. The BESIII collaboration has observed anomalously large BFs for the $D_s^+\to a_0(980)^{0(+)}\pi^{+(0)}$~\cite{BESIII:2019jjr}, $D_s^+\to a_0(980)^{0(+)}\rho^{+(0)}$~\cite{BESIII:2021aza}, and $D^{0(+)}\to a_0(980)^{+}\pi^{-(0)}$ decays~\cite{BESIII:2024tpv}. Particularly noteworthy is the measured ratio $\mathcal{B}(D^{+}\rightarrow a_{0}(980)^{+}\pi^{0})/\mathcal{B}(D^{+}\rightarrow a_{0}(980)^{0}\pi^{+}) = 2.6\pm0.6\pm 0.3$~\cite{BESIII:2024tpv}. 
This ratio is predicted to be much smaller than 1
in the $q\bar{q}$ model of $a_0(980)$, since $D\rightarrow a_{0}(980)^+P$ decays proceed primarily through external $W$-emission diagrams, which are suppressed by the extremely small coupling of the $W^+$ to the $a_0(980)^+$~\cite{Cheng:2024zul}. 
Examples of the external $W$-emission topological diagrams of $D^{+}\rightarrow a_{0}(980)^{+}\rho^{0}$ in the two-quark and tetraquark models are shown in Figs.~\ref{fig:feynman_diagram}(a) and \ref{fig:feynman_diagram}(b), respectively.
These observations provide compelling evidence for substantial FSI effecting the $D\to SP$ decays involving scalar mesons~\cite{Cheng:2024zul}.
Here $S$, $P$, and $V$ denote scalar, pseudo-scalar, and vector states, respectively.
Similarly, sizeable contributions from FSIs are anticipated to influence the $D^+ \to SV$ decays. The measurement of both the BFs for $D^+\to a_0(980)\rho$ decays and the relative ratio $\mathcal{B}(D^{+}\rightarrow a_{0}(980)^{+}\rho^{0})/\mathcal{B}(D^{+}\rightarrow a_{0}(980)^{0}\rho^{+})$ will offer crucial insights into the exotic structure of light scalar particles. 

Furthermore, although numerous $D_{(s)} \to SP$ decays have been studied~\cite{Boito:2008zk, Cheng:2022vbw, Dai:2023jix, Ikeno:2021kzf, Hsiao:2019ait, Cheng:2024zul, BESIII:2023htx}, the $D_{(s)}\to SS$ decay channel remains rarely explored experimentally and theoretically. In a tetraquark scenario,
the hadronization of each $S$ requires an additional pair of quark-antiquark. This makes
$D_{(s)}\to SS$ more sensitive to the role of FSIs in the production of scalar particles.
The intermediate process
$D^+ \to a_0(980)^+ f_0(500)$ of $D^+ \to \pi^+\pi^0\pi^0\eta$ and $D^+ \to \pi^+\pi^+\pi^-\eta$ is expected
to be governed by the external $W$-emission, and the BF should be suppressed by the smallness of the $a_0(980)^+$ decay constant under the assumption of $a_0(980)$ as a conventional two-quark meson in Fig.~\ref{fig:feynman_diagram}(a). However, this decay mode may receive contribution from other diagrams assuming $a_0(980)$ and $f_0(500)$ as tetraquark states. Examples are shown in Fig.~\ref{fig:feynman_diagram}(b), which could potentially enhance its BF significantly~\cite{Cheng:2024zul}.
The discovery of such decays would open a new window into the fundamental dynamics of charmed meson decays, particularly for probing exotic scalar meson configurations and their production mechanisms.

In addition, the $a_1(1260)$ is a low-lying axial vector meson. Its mass being close to the $\bar{K}K^*$ threshold implies the $K^*K$ channels may
play an essential role in its poorly understood internal structure. Some $D_{(s)}\to a_1(1260)P$ decays have been observed with large BFs in amplitude analyses of four-body $D_{(s)}$ decays~\cite{BESIII:2021aza,BESIII:2024muy,BESIII:2025nou, BESIII:2019lwn, BESIII:2019ymv}. 
We also search for $D^+\to a_1(1260)^+\eta$ in this Letter.

The $\psi(3770)$ decays predominantly into $D\bar{D}$ pairs without any additional hadrons. The excellent tracking, accurate calorimetry, and the large data sample at the $D\bar{D}$ threshold~\cite{BESIII:2009fln} provide an unprecedented opportunity to accurately study the intermediate processes of the $D^+ \to \pi^+ \pi^{+(0)} \pi^{-(0)} \eta$ decays. 
Based on a data set corresponding to an integrated luminosity of 20.3 fb$^{-1}$~\cite{Ablikim:2013ntc, BESIII:2024lbn}, this Letter reports the first amplitude analysis and updated BF
measurements of $D^+ \to \pi^+ \pi^{+(0)} \pi^{-(0)} \eta$.

A description of the design and performance of the BESIII detector can be found in Ref.~\cite{BESIII:2009fln}. Simulated data samples are produced with a {\sc geant4}-based~\cite{GEANT4:2002zbu} software package, which includes the geometric description~\cite{Huang:2022wuo} of the BESIII detector and the detector response. The simulation models the beam energy spread and initial state radiation (ISR) in the $e^+e^-$ annihilations with the generator {\sc kkmc}~\cite{Jadach:2000ir}. The inclusive Monte Carlo (MC) sample includes the production of $D\bar{D}$ pairs, the non-$D\bar{D}$ decays of the $\psi(3770)$, the ISR production of the $J/\psi$ and $\psi(3686)$ states, and the continuum processes incorporated in {\sc kkmc}. All particle decays are modeled with {\sc evtgen}~\cite{Lange:2001uf, Ping:2008zz} using BFs either taken from the Particle Data Group~(PDG)~\cite{PDG}, when available, or otherwise estimated with {\sc lundcharm}~\cite{Chen:2000tv, Yang:2014vra}. Final state radiation from charged final state particles is incorporated using {\sc photos}~\cite{Richter-Was:1992hxq}.
Charge-conjugate modes are implied throughout this Letter.

The double-tag~(DT) method~\cite{MARK-III:1985hbd} provides high-purity samples for measuring the absolute BFs of $D$ meson decays, eliminating dependence on the integrated luminosity and the $D\bar{D}$ production cross section~\cite{BESIII:2024zvp}. The tagged $D^-$ mesons are first reconstructed in one of the five decay modes $K^+ \pi^- \pi^-$, $K^+ \pi^- \pi^- \pi^0$, $K_{S}^0 \pi^-$, $K_{S}^0 \pi^- \pi^0$, and  $K_{S}^0 \pi^- \pi^- \pi^+$, referred to as single-tag~(ST) events. 
An event in which both a signal $D^+ \to \pi^+ \pi^{+(0)} \pi^{-(0)} \eta$ decay and a tagged $D^-$ is simultaneously found is referred to as a DT event.
The selection criteria of the final state particles
are the same as Ref.~\cite{BESIII:2024ncc}.


The $D$ mesons are identified using the energy difference $\Delta E \equiv  E_{D}-E_{\rm beam}$ and the beam-constrained mass $M_{\rm BC} \equiv \sqrt{E^{2}_{\rm beam}/c^{4}-|\vec{p}_{\bar D}|^{2}/c^{2}}$, where $E_{\rm beam}$ is the beam energy, and $\vec{p}_{D}$ and $E_{D}$ are the  measured momentum and energy of the $D$ candidate in the $e^+e^-$ rest frame, respectively.
The $M_{\rm BC}$ and $\Delta E$ distributions of the $D^{\pm}$ signals are expected to be around the known $D^{\pm}$ mass~\cite{PDG} and zero, respectively.
The tagged $D^-$ is reconstructed through the tag modes.
In case of multiple candidates, the one with the minimum $|\Delta{E}|$ is chosen. More details about the ST information can be found in Ref.~\cite{BESIII:2025lyg}.

Once a ST $D^-$ decay has been identified, a search is made for a $D^+$ signal decay in the same event. The signal candidates are reconstructed using the remaining charged tracks and neutral showers. 
The best signal candidate with the minimum $|\Delta{E}|$ is selected.
All $D^{\pm}$ candidates are required to satisfy $1.863 < M_{\rm BC} < 1.877$ GeV/$c^2$, and $-0.033~(-0.052) < \Delta E < 0.032~(0.038)$ GeV for $\pi^+ \pi^{+(0)} \pi^{-(0)} \eta$.
For the $\pi^{+}\pi^{-}[\pi^{0}\pi^{0}]$ and $\pi^{+(0)}\pi^{-(0)}\eta$ combinations, the $K_{S}^{0}$ and $\eta^{\prime}$ contributions are rejected by requiring their invariant masses to be outside $(0.478,0.517)~\mathrm{GeV}/c^{2}$ $[(0.448,0.534)~\mathrm{GeV}/c^{2}]$ and $(0.8,1.0)~\mathrm{GeV}/c^{2}$, respectively. These requirements correspond to at least five times the fitted mass resolution away from the individual nominal mass. 
For $\pi^+ \pi^{0} \pi^{0} \eta$ decays, the peaking backgrounds from $\pi^+ \pi^{0} \pi^{0} \pi^{0}$ are rejected by requiring no $\pi^+ \pi^{0} \pi^{0} \pi^{0}$ combinations in the same candidate event that satisfy both $\Delta E_{\pi\pi^0\pi^0\pi^0} \in (-0.1, 0.1)~\mathrm{GeV}$ and $M_{\rm BC}^{\pi\pi^0\pi^0\pi^0} \in (1.83, 1.89)~\mathrm{GeV}/c^{2}$. This combined selection rejects more than 77\% of the background while retaining 98\% of the signal efficiency~\cite{BESIII:2020pxp}.
Furthermore, a kinematic fit is performed with some constraints: 
(1) the four-momenta of final-state particles are constrained to match the initial $e^+e^-$ system momentum, 
and (2) the invariant masses of intermediate states ($K_S^0$, $\pi^0$, and tagged $D^-$ on the tag side; $\pi^0$ and $\eta$ on the signal side) are constrained to their known values~\cite{PDG}. The requirements $\chi^2<100(50)$ are applied for $\pi^+ \pi^{+(0)} \pi^{-(0)} \eta$.
Finally, for $\pi^+ \pi^{+(0)} \pi^{-(0)} \eta$ channels, there are 3122 (648) DT events obtained for the amplitude analysis with a signal purity of $(89.3\pm0.5)$\% ($(78.0\pm 1.4)$\%)~\cite{supplement}.
The amplitude analysis requires a sample with good resolution and all candidates falling within the phase-space boundary. Therefore, a subsequent kinematic fit is performed, which incorporates the previous constraints and introduces an additional constraint on the invariant mass of the signal $D^+$ meson. 
The amplitude analysis is then conducted using the four-momenta of particles obtained from this kinematic fit.

An unbinned maximum likelihood method is adopted in the amplitude analysis of the $D^{+} \to \pi^+ \pi^{+(0)} \pi^{-(0)} \eta$ decays. The likelihood function is constructed with a probability density function~(PDF) in which the momenta of the four final-state particles are used as inputs. The total likelihood is the product of the likelihoods for all events in the data samples. The total amplitude is modeled as a coherent sum over all intermediate processes $M(p_{j})=\sum\rho_{n}e^{i\phi_{n}}A_{n}(p_{j})$, where $\rho_{n}e^{i\phi_{n}}$ is the coefficient of the $n^{\mathrm{th}}$ amplitude with magnitude $\rho_{n}$ and phase $\phi_{n}$. 
The amplitude is symmetrized under the exchange of the two identical $\pi^{+(0)}$ mesons to incorporate the Bose symmetry.
The $n^{\mathrm{th}}$ amplitude $A_{n}(p_{j})$ is given by $A_{n}=P_{n}^{1}P_{n}^{2}S_{n}F_{n}^{1}F_{n}^{2}F_{n}^{3}$, where the indices 1, 2 and 3 correspond to the two subsequent intermediate resonances and the $D^{+}$ meson, respectively, $F_{n}^{i}$ is the Blatt-Weisskopf barrier factor~\cite{Zou:2002ar} and $P_{n}^{i}$ is the propagator of the intermediate resonance. The function $S_{n}$ describes the angular momentum conservation in the decay and is constructed using the covariant tensor formalism~\cite{Zou:2002ar}.
The relativistic Breit-Wigner~(RBW)~\cite{Jackson:1964zd} function is used to describe the propagator for the resonances $\eta(1405)$, $f_{1}(1285)$ and $f_{1}(1420)$. The resonance $\rho(770)$ is parameterized by the Gounaris-Sakurai~\cite{Gounaris:1968mw} lineshape, and the $a_{0}(980)$ is parameterized by a coupled Flatt\'{e} formula, and the parameters are fixed to the values given in Ref.~\cite{Abele:1998qd}. We use the same parameterization to describe $f_{0}(500)$ as in Ref.~\cite{Bugg:1996ki}. The masses and widths of the intermediate resonances, except for $a_{0}(980)$ and $f_{0}(500)$, are taken from the PDG~\cite{PDG}.

The background PDF, $B(p_{j})$, which is derived from the inclusive MC sample using the XGBoost package~\cite{Rogozhnikov:2016bdp,Liu:2019huh}, is then added incoherently to the signal PDF.
The details can be found in the supplemental material~\cite{supplement}.
As a consequence, the combined PDF can be written as~\cite{BESIII:2021aza}
\begin{equation}
\epsilon R_{4}\left[f_{s}\frac{|M(p_{j})|^{2}}{\int\epsilon|M(p_{j})|^{2}R_{4}dp_{j}}+(1-f_{s})\frac{B_{\epsilon}(p_{j})}{\int\epsilon B_{\epsilon}(p_{j})R_{4}dp_{j}}\right],
\label{eq: the combined PDF}
\end{equation}
where $\epsilon$ is the acceptance function, $B_{\epsilon}(p_{j}) = B(p_{j})/\epsilon$, $f_s$ is the signal purity, 
and $R_{4}dp_{j}$ is the element of the four-body phase space. The normalization integral in the denominator is determined by an MC technique as described in Ref.~\cite{CLEO:2012beo}.

In the data projections, the decay channels $D^{+} \to \pi^+ \pi^{+(0)} \pi^{-(0)} \eta$ exhibit distinct resonance structures: the former shows prominent $a_{0}(980)^{+}$ and $\rho^{0}$ signals, while the latter displays clear $a_{0}(980)^{0}$ and $\rho^{+}$ peaks.
Therefore, the decays $D^{+} \to a_{0}(980)^{+} \rho^{0}$ and $D^{+} \to a_{0}(980)^{0} \rho^{+}$ are chosen as the reference amplitudes for the $\pi^+\pi^+\pi^-\eta$ and $\pi^+\pi^0\pi^0\eta$ channels, respectively. 
We consider some possible amplitudes involving the resonances $f_{0}(500)$, $f_{0}(980)$, $f_{1}(1285)$, $\eta(1295)$, $\eta(1405)$, $\eta(1475)$, $f_{1}(1420)$ and $f_{1}(1510)$, as well as other non-resonant components.
All contributions with significance larger than $5\sigma$ are retained for further analysis.
Here, the significance is calculated using the changes of $\ln L$ and the number of degrees of freedom when the fit is performed with and without including the corresponding amplitudes. Following this criterion, six intermediate states are retained in the fit model for both channels. 
Amplitudes are symmetrized with respect to the two identical $\pi^{+(0)}$ mesons and charge-conjugation constraints through Clebsch-Gordan coefficients (see the supplemental material~\cite{supplement} for details).

\begin{figure}[htbp]
  \centering
  \hspace*{-2em}
  \includegraphics[width=0.48\textwidth]{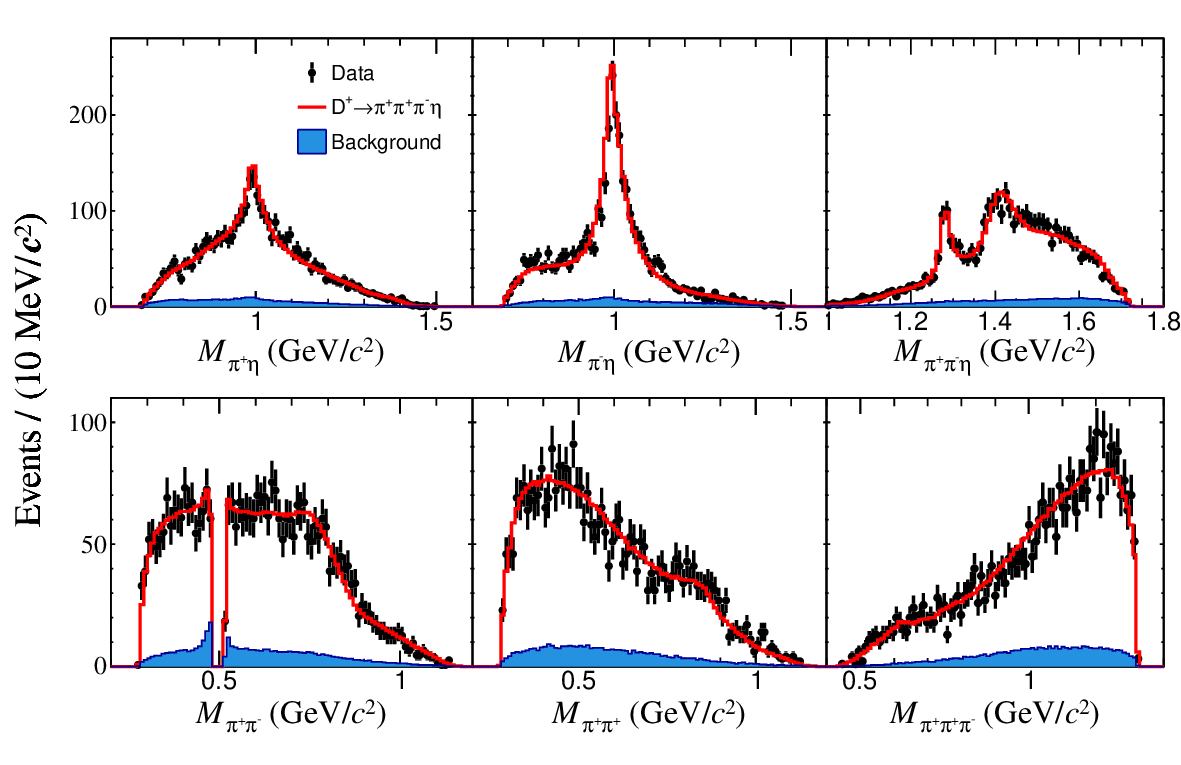}
  
  \hspace*{-2em}
  \includegraphics[width=0.48\textwidth]{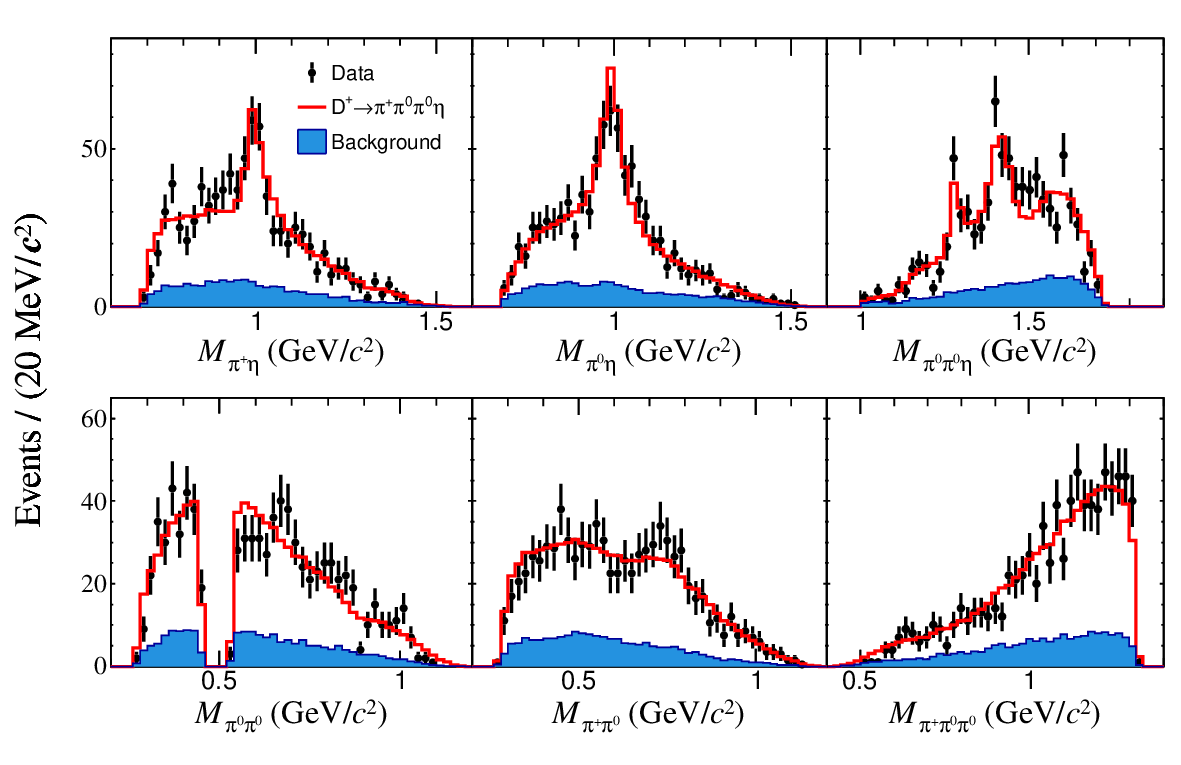}
  \caption{Mass projections of the fit results. The plots containing identical $\pi^+$ or $\pi^0$ are added into one projection. 
  The data are represented by points with error bars and the fit results by the red curves.
  The blue curves indicate the background contribution estimated with inclusive MC samples. 
  }
  \label{fig:pwa_fitresult_mass}
\end{figure}

The decay amplitudes and the corresponding $\phi_{n}$, fit fraction~(FF$_{n}$) and BF$_{n}$ are listed in Table~\ref{tab:phase, FF, BF}. The mass projections of the fit are shown in Fig.~\ref{fig:pwa_fitresult_mass}.
For the $n^{\text{th}}$ component, its contribution relative to the total BF is quantified by
$\mathrm{FF}_{n} = \int |\rho_{n}A_{n}(p_{j})|^{2}R_{4}\,dp_{j}/\int |M|^{2}R_{4}\,dp_{j}$.
We evaluate the goodness-of-fit using the mixed-sample method~\cite{Williams:2010vh,BESIII:2019lwn}. This method quantifies the consistency between the data and the signal MC in the multi-dimensional phase space. A test statistic $T$ is calculated based on the fraction of nearest neighbours for each event that originate from the other sample. The pull of $T$ from its expected value under the hypothesis of a perfect fit is then computed.  For the $\pi^+ \pi^{+(0)} \pi^{-(0)} \eta$ channel, we find a pull value of 0.51 (0.57), demonstrating the good fit quality.


\begin{table*}[htbp]
    \caption{Phase~($\phi$), fit fraction~(FF), 
    statistical significance and BF of each amplitude. 
    The first and second uncertainties are statistical and systematic, respectively. The intermediate states are reconstructed in the decays $a_0(980) \to \eta \pi$, $f_0(500)\to\pi \pi$, $\rho(770) \to \pi \pi$ and $\eta(1405)/f_1(1285)/f_1(1420) \to a_0(980) \pi$.
    }
    \label{tab:phase, FF, BF}
    \begin{center}
    \renewcommand{\arraystretch}{1.16}
    \setlength{\tabcolsep}{12pt} 
    \begin{tabular*}{0.95\linewidth} {l @{\extracolsep{\fill}} c c c c}
    \hline \hline 
    Amplitude & $\phi$ & FF(\%) & Significance~($\sigma$)  & BF($\times 10^{-4}$) \\
    \hline
    
    \multicolumn{5}{l}{$D^+ \to \pi^+ \pi^+ \pi^- \eta$ decay:} \\
    
$D^+ \to a_0(980)^+ \rho(770)^0$ & 0~(fixed) & $14.7 \pm 1.1 \pm 1.0$ & $>10$ & $4.7 \pm 0.4 \pm 0.3$ \\$D^{+}\to a_0(980)^{+}f_0(500)$ & $5.5 \pm 0.1 \pm 0.2$ & $23.2 \pm 2.3 \pm 2.2$ & $>10$ & $7.5 \pm 0.7 \pm 0.7$ \\$D^+\to \eta(1405)\pi^+$ & $2.4 \pm 0.1 \pm 0.1$ & $11.2 \pm 1.4 \pm 1.5$ & $>10$ & $3.6 \pm 0.5 \pm 0.5$ \\$D^+\to f_1(1285)\pi^+$ & $4.9 \pm 0.1 \pm 0.3$ & $17.3 \pm 0.9 \pm 0.8$ & $>10$ & $5.5 \pm 0.3 \pm 0.3$ \\$D^+\to f_1(1420)\pi^+$ & $6.2 \pm 0.1 \pm 0.1$ & $16.4 \pm 1.6 \pm 2.5$ & $>10$ & $5.2 \pm 0.5 \pm 0.8$ \\$D^+\to [a_0(980)^\pm \pi^\mp]_{S-wave}\pi^+$ & $3.5 \pm 0.1 \pm 0.3$ & $34.3 \pm 2.8 \pm 3.2$ & $>10$ & $11.0 \pm 0.9 \pm 1.0$ \\

    \hline

    \multicolumn{5}{l}{$D^+ \to \pi^+ \pi^0 \pi^0 \eta$ decay:} \\
    
$D^+ \to a_0(980)^0 \rho(770)^+$ & 0~(fixed) & $35.0 \pm 3.5 \pm 2.6$ & $>10$ & $8.5 \pm 0.9 \pm 0.6$ \\$D^+ \to a_0(980)^+ f_0(500)$ & $0.4 \pm 0.1 \pm 0.0$ & $34.6 \pm 3.1 \pm 1.0$ & $>10$ & $8.4 \pm 0.8 \pm 0.3$ \\$D^+ \to \eta(1405)\pi^+$ & $2.6 \pm 0.3 \pm 0.2$ & $4.3 \pm 1.7 \pm 0.7$ & 5.1 & $1.0 \pm 0.4 \pm 0.2$ \\$D^+ \to f_1(1285)\pi^+$ & $2.9 \pm 0.2 \pm 0.3$ & $12.8 \pm 1.9 \pm 0.7$ & $>10$ & $3.1 \pm 0.5 \pm 0.2$ \\$D^+ \to f_1(1420)\pi^+$ & $5.9 \pm 0.2 \pm 0.3$ & $17.8 \pm 3.0 \pm 1.6$ & $>10$ & $4.3 \pm 0.8 \pm 0.4$ \\$D^+ \to [a_0(980)^0\pi^0]_{S-wave}\pi^+$ & $4.4 \pm 0.2 \pm 0.2$ & $11.0 \pm 2.6 \pm 0.9$ & $8.6$ & $2.7 \pm 0.6 \pm 0.2$ \\

    \hline \hline
    \end{tabular*}
    \end{center}
\end{table*}

The systematic uncertainties in the amplitude analysis arise from the following sources:  
(I)~the amplitude model, including the mass and coupling constants of the $a_{0}(980)^{\pm(0)}$, which are varied within the uncertainties reported in Ref.~\cite{Abele:1998qd}, the masses and widths of the $\eta(1405)$, $f_{1}(1285)$, and $f_{1}(1420)$ states, varied according to the uncertainties listed in the PDG~\cite{PDG}, and the $f_{0}(500)$ line shape, which is altered by using an alternative RBW function with its mass and width fixed to 526~MeV/$c^{2}$ and 535~MeV, respectively, as in Ref.~\cite{BES:2004mws};
(II)~the effective barrier radii for $D^{+}$ and other intermediate states~(R value), estimated by varying the nominal values by $\pm 1$~GeV${}^{-1}$;
(III)~experimental effects related to the detector response, evaluated using the same method as described in Ref.~\cite{BESIII:2020ctr}; 
(IV)~the fit bias, assessed by fitting 600 signal MC samples of the same size as the data sample: the results show good agreement between the fitted and input parameters of the amplitude model, with no significant bias observed; 
(V)~the background estimation: the uncertainty associated with the background sample size has been studied by varying the purity within its statistical uncertainty, while the uncertainty from the peaking background~\cite{supplement} has been assessed by using an alternative sample with their fractions varied within the BF uncertainties.
These systematic uncertainties are estimated separately by taking the difference between the values of $\phi_{n}$ and FF${}_{n}$  obtained by the alternative and the baseline fits. The total systematic uncertainties are obtained by adding each term in quadrature, and the details are listed in the supplemental material~\cite{supplement}.

Furthermore, we measure the BFs of $D^{+} \to \pi^+ \pi^{+(0)} \pi^{-(0)} \eta$ with the DT technique~\cite{MARK-III:1985hbd}, using the same five tag modes and selection criteria as in the amplitude analysis. 
The BFs of the signal decays are determined by 
\begin{equation}
  \mathcal{B}=\frac{N_{\rm DT}}{\sum_{\alpha}({N^{\alpha}_{\rm ST}}\epsilon^{\alpha}_{\rm DT}/\epsilon^{\alpha}_{\rm ST})\mathcal{B}_{\rm sub}}\,,
  \label{eq:bf}
\end{equation}
where $\alpha$ denotes the tag mode, $N^{\alpha}_{\rm ST}$ and $N_{\rm DT}$ are the ST and total DT yields, respectively, and $\mathcal{B}_{\rm sub}$ is the product of the BFs of all intermediate particles in the decay chain. The efficiencies $\epsilon^{\alpha}_{\rm ST}$ and $\epsilon^{\alpha}_{\rm DT}$ are obtained from MC simulation, with $\epsilon^{\alpha}_{\rm DT}$ evaluated using the amplitude model of the signal decays. All quantities—$N^{\alpha}_{\rm ST}$, $\epsilon^{\alpha}_{\rm ST}$, and $\epsilon^{\alpha}_{\rm DT}$—are detailed in the supplemental material~\cite{supplement}, where the 2D fit procedure used to extract $N_{\rm DT}$ is also described.
Finally, we obtain the absolute BFs with the highest precision to date:  
$\mathcal{B}(D^+ \to \pi^+ \pi^+ \pi^- \eta |_{ \rm non- \eta^\prime})=(3.20\pm0.06_{\text{stat.}}\pm0.03_{\text{syst.}})\times 10^{-3}$ and $\mathcal{B}(D^+ \to \pi^+ \pi^0 \pi^0 \eta |_{ \rm non- \eta^\prime})=(2.43 \pm 0.11_{\text{stat.}} \pm 0.04_{\text{syst.}}) \times 10^{-3}$, which are consistent with the previous measurements~\cite{BESIII:2020pxp}.


The systematic uncertainties in the BF measurement are summarized in the supplemental material~\cite{supplement} and discussed below.
The uncertainty in the ST $D^-$ yield reflects the uncertainty in the fit to the $M_{\rm BC}$ distributions and is assessed by varying the signal and background shapes.
The uncertainties in the tracking or PID of $\pi^{\pm}$ are studied with control samples of DT-hadronic events.
The uncertainty associated with the $\eta$ and $\pi^0$ reconstruction are both assigned by studying a control sample of $D^0 \to K^- \pi^+ \pi^0$ decays.
The possible difference between data and MC simulation is accounted for by examining the $\Delta E_{\text{sig}}$ cut efficiency after applying a double-Gaussian smearing to the signal MC sample. The observed efficiency variation is taken as the systematic uncertainty.
The uncertainty from the amplitude model is determined by varying the amplitude model parameters based on their error matrix.
For the 2D fit, the mean and width of the convolved Gaussian function are varied by $\pm 1\sigma$ for the signal shape; the ARGUS end-point for the background shape is varied by $\pm$0.2 MeV; the fixed yields of peaking background decays are varied by $\pm 1\sigma$ according to their quoted BFs~\cite{PDG,supplement}.
The uncertainty due to $\mathcal{B}(\pi^0/\eta \to \gamma \gamma)$ is taken from the PDG~\cite{PDG}.
The effect of the limited MC sample size is also included.

In summary, using an $e^+e^-$ collision data sample corresponding to an integrated luminosity of 20.3~$\rm{fb}^{-1}$ collected by the BESIII detector at $\sqrt{s}=3.773$~GeV, the amplitude analysis of the singly Cabibbo-suppressed decays $D^+ \to \pi^+ \pi^{+(0)} \pi^{-(0)} \eta$ is performed for the first time, with the absolute BFs achieving threefold improvement in precision.
The absolute BFs of intermediate states are calculated as $\mathcal{B}_{i}=\text{FF}_{i}\times\mathcal{B}(D^+ \to \pi^+ \pi^{+(0)} \pi^{-(0)} \eta)$ and listed in Table~\ref{tab:phase, FF, BF}. 
The $D\to SS$ process $D^{+}\to a_0(980)^{+}f_0(500)$ has been observed in both two channels for the first time, with a significance exceeding $10\sigma$. The measured BF reaches an unexpectedly large level of $10^{-3}$, which strongly supports the tetraquark interpretation of light scalar particles within the topological diagrams frameworks \cite{Cheng:2024zul,Chau:1986jb,Cheng:2010ry}.

Moreover, in the $D\to SV$ sector, we report the first observation of both $D^+ \to a_0(980)^+ \rho(770)^0$ and $D^+ \to a_0(980)^0 \rho(770)^+$ decays with statistical significances exceeding $10\sigma$. Based on these results, the ratio 
$\frac{\mathcal{B}(D^+ \to a_0(980)^+ \rho(770)^0)}{\mathcal{B}(D^+ \to a_0(980)^0 \rho(770)^+)}$
is first measured to be $0.55\pm0.08_{\text{stat.}}\pm0.05_{\text{syst.}}$. 
Our measured ratio, together with the earlier BESIII result in $D \to a_0 \pi$ decays~\cite{BESIII:2024tpv}, implies substantial FSI effects, and provide a new perspective on the nature of the $a_0(980)$ states.

Notably, no significant $a_1(1260)\eta$ signal is observed, which contradicts with the abundant $a_1(1260)$ typically seen in other $D\to P 3\pi$ decays~\cite{BESIII:2023qgj,BESIII:2023exz}, suggesting anomalous dynamics in this specific final state.




The BESIII Collaboration thanks the staff of BEPCII (https://cstr.cn/31109.02.BEPC) and the IHEP computing center for their strong support. This work is supported in part by National Key R\&D Program of China under Contracts Nos. 2023YFA1606000, 2023YFA1606704; National Natural Science Foundation of China (NSFC) under Contracts Nos. 11635010, 11935015, 11935016, 11935018, 12025502, 12035009, 12035013, 12061131003, 12192260, 12192261, 12192262, 12192263, 12192264, 12192265, 12221005, 12225509, 12235017, 12342502, 12361141819; the Chinese Academy of Sciences (CAS) Large-Scale Scientific Facility Program; the Strategic Priority Research Program of Chinese Academy of Sciences under Contract No. XDA0480600; CAS under Contract No. YSBR-101; 100 Talents Program of CAS; The Institute of Nuclear and Particle Physics (INPAC) and Shanghai Key Laboratory for Particle Physics and Cosmology; ERC under Contract No. 758462; German Research Foundation DFG under Contract No. FOR5327; Istituto Nazionale di Fisica Nucleare, Italy; Knut and Alice Wallenberg Foundation under Contracts Nos. 2021.0174, 2021.0299, 2023.0315; Ministry of Development of Turkey under Contract No. DPT2006K-120470; National Research Foundation of Korea under Contract No. NRF-2022R1A2C1092335; National Science and Technology fund of Mongolia; Polish National Science Centre under Contract No. 2024/53/B/ST2/00975; STFC (United Kingdom); Swedish Research Council under Contract No. 2019.04595; U. S. Department of Energy under Contract No. DE-FG02-05ER41374

\appendix

\section{I. two-dimensional fit to the distribution of $M_{\rm BC}^{\rm tag}$ versus $M_{\rm BC}^{\rm sig}$}

The same selection criteria are applied as those used in the amplitude analysis and branching fraction~(BF) measurement. 
After applying all criteria discussed  in the main text, the signal yield is extracted from a two-dimensional~(2D) binned maximum likelihood fit on the distribution of $M_{\rm BC}^{\rm tag}$~($y$) versus $M_{\rm BC}^{\rm sig}$~($x$).
Signal events with both tag and signal sides reconstructed correctly are expected to  concentrate around $M_{\rm{BC}}^{\rm{sig}} = M_{\rm{BC}}^{\rm{tag}} = M_{D^+}$, where $M_{D^+}$ is the known $D^+$ mass~\cite{PDG}. Besides signal events, we define four kinds of background events. Candidates with correctly reconstructed signal side $D^+$~(or tag side $D^-$) and incorrectly reconstructed tag side $D^-$~(or signal side $D^+$) are BKGI, which appear around the lines $M_{\rm{BC}}^{\rm{sig}}$ or $M_{\rm{BC}}^{\rm{tag}}$ = $M_{D^+}$. Other candidates appearing around the diagonal are mainly from the $e^+e^-\to D^0\bar{D}^0$ mispartition and the $e^+e^- \to q\bar{q}$ processes (BKGII). 
The remaining combinatorial backgrounds come mainly from candidates reconstructed incorrectly on both sides (BKGIII). For $D^+\to\pi^+\pi^+\pi^-\eta$, there are two types of peaking backgrounds~($D^+ \to K_S^0(\to \pi^+ \pi^-) \pi^+ \eta$ 
, $D^+ \to \eta^\prime(\to \pi^+ \pi^- \eta) \pi^+$),
indicated as BKGIV. 
The peaking backgrounds for $D^+\to\pi^+\pi^0\pi^0\eta$ include the following process (the numbers of each component represent their proportions with respect to the total peaking background):

\begin{itemize}
    \item $D^+ \to K_S^0(\to \pi^0 \pi^0) \pi^+ \eta$~(23.2\%): this background originates from the $K_S^0$ peak, where the $K_S^0$ decays into two $\pi^0$ mesons;
    \item $D^+ \to K_S^0(\to \pi^0 \pi^0) \pi^+ \pi^0$~(32.5\%) and $D^+ \to \pi^+ \pi^0 \pi^0 \pi^0$~(12.7\%): these backgrounds arise primarily from the misidentification of $\pi^0$ mesons as $\eta$ mesons;
    \item $D^+ \to K_S^0(\to \pi^0 \pi^0) \pi^+ \pi^0 \pi^0$~(26.8\%) and $D^+ \to \pi^+ \pi^0 \pi^0 \pi^0 \pi^0$~(4.8\%): these backgrounds are mainly due to the misidentification of $\pi^0$ mesons as $\eta$ mesons, as well as the presence of additional $\pi^0$ mesons in the final state.
\end{itemize}

\noindent\hrulefill

The probability density functions~(PDFs) used in the 2D fit for different components are:
\begin{itemize}
\item Signal: $s(x, y)$,
\item BKGI: $b_1(x) \cdot  {\rm ARGUS}(y; m_0, c, p) + b_2(y) \cdot {\rm ARGUS}(x; m_0, c, p)$,
\item BKGII: ${\rm ARGUS}((x+y)/\sqrt{2}; m_0, c, p) \cdot g((x-y)/\sqrt{2})$,
\item BKGIII: ${\rm ARGUS}(x; m_0, c, p) \cdot {\rm ARGUS}(y; m_0, c, p)$,
\item BKGIV: $b_p(x, y)$.
 \end{itemize}
The signal shape $s(x, y)$ is described by the 2D MC-simulated shape convolved with a 2D Gaussian. The parameters of the Gaussian function are obtained by the 1D fit on $M_{\rm{BC}}$ of the signal and tag sides, separately, and are fixed in the 2D fit. For BKGI, $b_{1,2}(x, y)$ is described by the 1D MC-simulated shape convolved with a Gaussian function, ARGUS($x, y$) is the ARGUS function. For BKGII, it is described by an ARGUS function in the diagonal axis multiplied by a Gaussian function in the anti-diagonal axis.  For BKGIII, it is  construted by an ARGUS function in $M_{\rm{BC}}^{\rm{sig}}$ multiplied by an ARGUS function in $M_{\rm{BC}}^{\rm{tag}}$. In the fit, the parameters $m_0$ and $p$ for ARGUS are fixed at 1.8865 GeV/$c^2$ and 0.5, respectively. For peaking backgrounds, the $b_p(x, y)$ shapes are extracted from MC simulation.  
Specifically, we select the corresponding peaking backgrounds from the inclusive MC sample.
Finally, the peaking background yields are fixed according to the inclusive MC sample. 

\begin{figure*}[t]  
    \centering
    \includegraphics[width=1.0\textwidth]{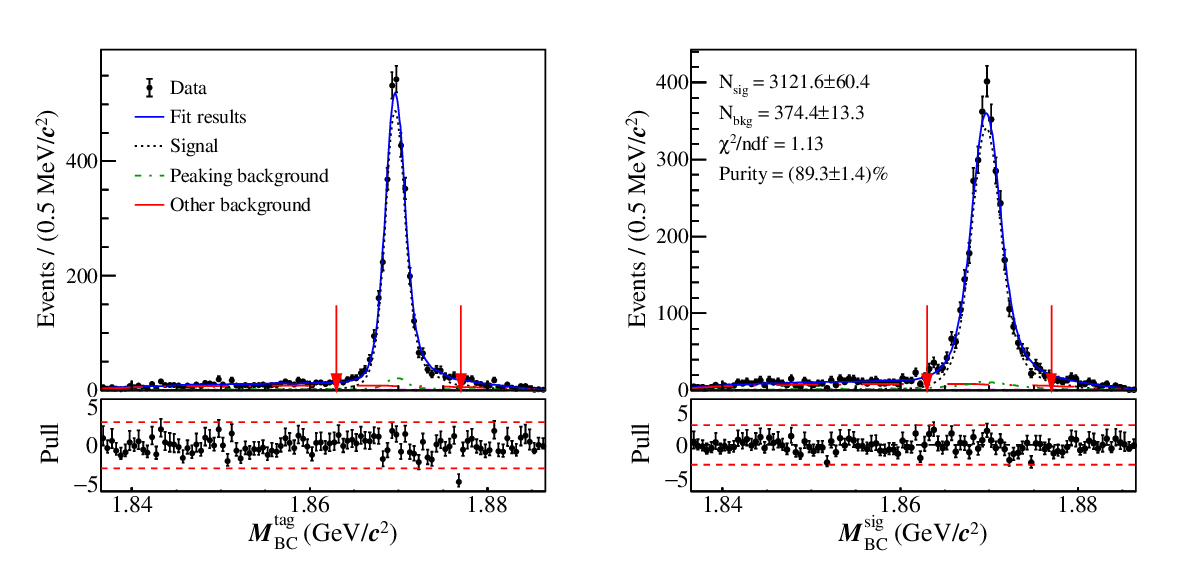}
    
    
    \caption{Projections on $M_{\rm BC}^{\rm tag}$~(a) and $M_{\rm BC}^{\rm sig}$~(b) distributions of the 2D fit for $D^+\to\pi^+\pi^+\pi^-\eta$. The points with error bars are data, the blue solid curves are the total fits, the black dotted curves are the signal shape, the green dash-dotted curves are the peaking background shapes, and the red long-dashed-dotted curves are other background shapes. 
    The red arrows mark the signal region. 
    The red dashed lines denote the $\pm3$ standard deviations in the pull distributions.
    }
    \label{fig:2Dfit_nominal_for_DpToEta3Pi} 
\end{figure*}

\begin{figure*}[t]  
    \centering
    \includegraphics[width=1.0\textwidth]{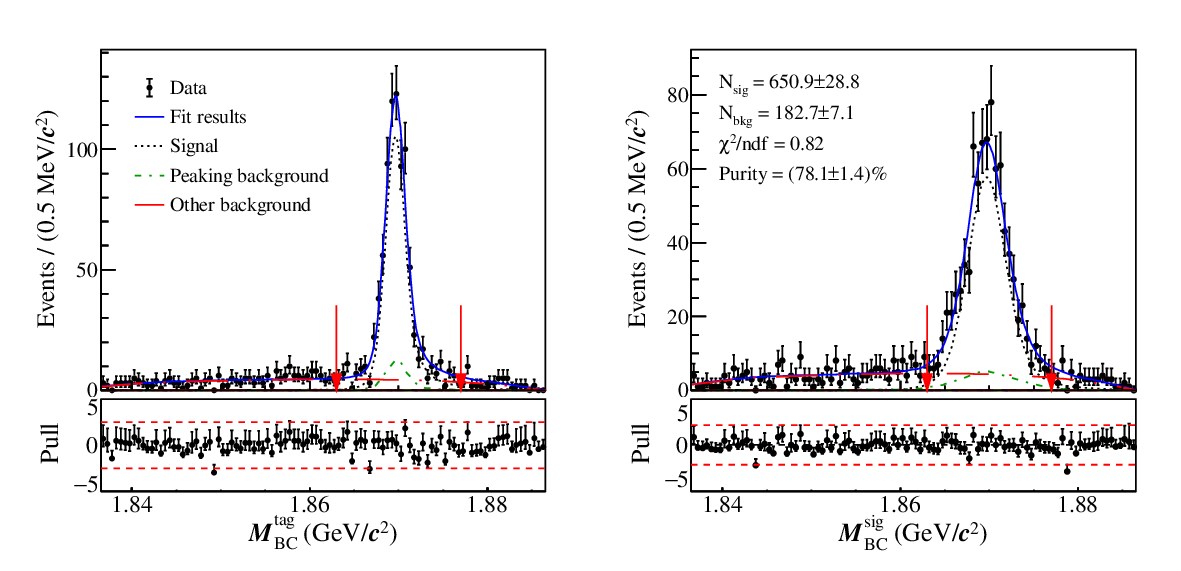}
    
    
    \caption{Projections on $M_{\rm BC}^{\rm tag}$~(a) and $M_{\rm BC}^{\rm sig}$~(b) distributions of the 2D fit for $D^+\to\pi^+\pi^0\pi^0\eta$. The points with error bars are data, the blue solid curves are the total fits, the black dotted curves are the signal shape, the green dash-dotted curves are the peaking background shapes, and the red long-dashed-dotted curves are other background shapes. 
    The red arrows mark the signal region. 
    The red dashed lines denote the $\pm3$ standard deviations in the pull distributions.
    }
    \label{fig:2Dfit_nominal_for_DpToEta2Pi0Pi}
\end{figure*}

\begin{figure*}[t]  
    \centering
    \begin{overpic}[width=0.49\textwidth]{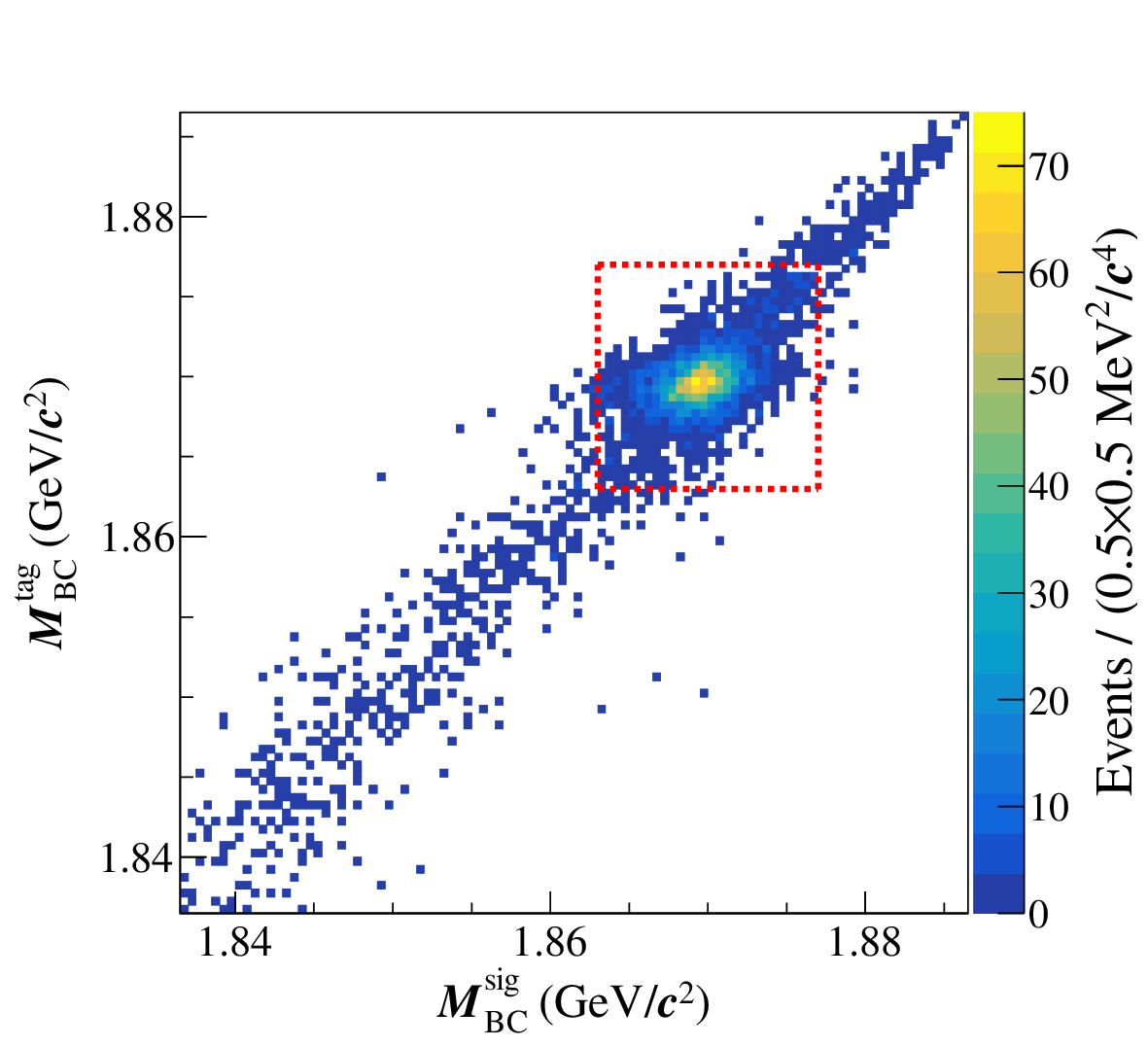}
    \end{overpic}%
    \hfill 
    \begin{overpic}[width=0.49\textwidth]{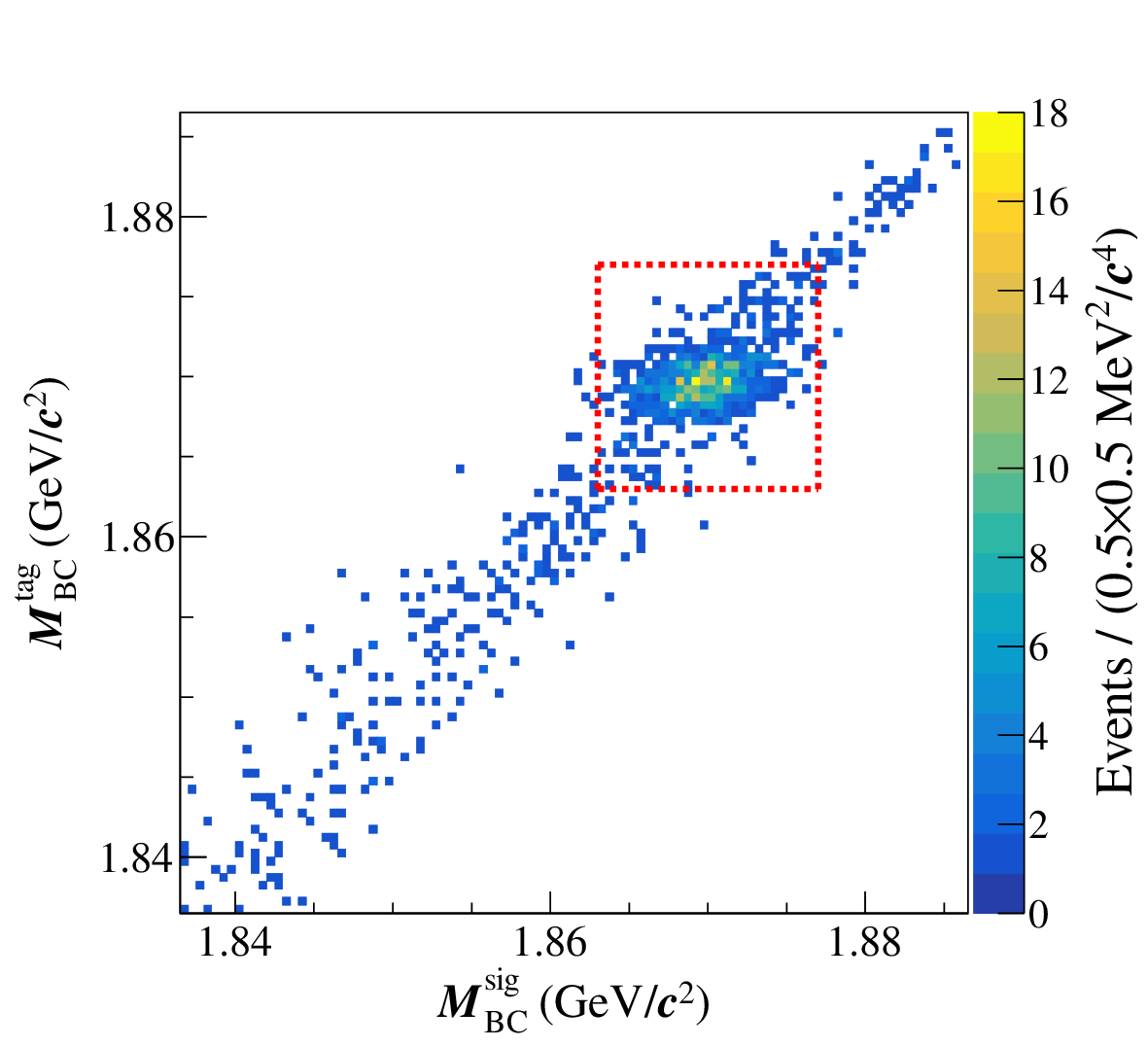}
    \end{overpic}
    \caption{The 2D distributions of $M_{\rm BC}^{\rm tag}$ versus $M_{\rm BC}^{\rm sig}$ for $D^+\to\pi^+\pi^+\pi^-\eta$ (left) and $D^+\to\pi^+\pi^0\pi^0\eta$ (right), with the signal regions marked by red boxes.}
    \label{fig:MBC_tag_and_MBC_sig_2D_distribution}
\end{figure*}

Finally, the projections of the 2D fit on the $M_{\rm BC}^{\rm tag}$ and $M_{\rm BC}^{\rm sig}$ distributions for $D^+ \to \pi^+ \pi^+ \pi^- \eta$ and $D^+ \to \pi^+ \pi^0 \pi^0 \eta$ are shown in Figs.~\ref{fig:2Dfit_nominal_for_DpToEta3Pi} and \ref{fig:2Dfit_nominal_for_DpToEta2Pi0Pi}, respectively. Additionally, the 2D distributions of $M_{\rm BC}^{\rm tag}$ versus $M_{\rm BC}^{\rm sig}$ for data in these two decay channels are presented in Fig.~\ref{fig:MBC_tag_and_MBC_sig_2D_distribution}.

\section{II. The Background PDF modeling in amplitude analysis}

The background PDF $B(p_{j})$ is modeled using the XGBoost~\cite{Rogozhnikov:2016bdp,Liu:2019huh} package by training a binary classifier to distinguish between background events from the inclusive MC sample and phase-space (PHSP) MC events within the signal region. 
Since the two positive pions in the final state of $D^+ \to \pi^+ \pi^+ \pi^- \eta$ are identical, they are not sorted and are labeled as $\pi_1^+$ and $\pi_2^+$ randomly. The input features consist of ten invariant-mass-squared variables: $M_{\pi_1^+\pi_2^+}$, $M_{\pi_1^+\pi^-}$, $M_{\pi_2^+\pi^-}$, $M_{\pi_1^+\pi_2^+\pi^-}$, $M_{\pi_1^+\eta}$, $M_{\pi_2^+\eta}$, $M_{\pi^-\eta}$, $M_{\pi_1^+\pi^-\eta}$, $M_{\pi_2^+\pi^-\eta}$, and $M_{\pi_1^+\pi_2^+\eta}$. Similarly, for $D^+ \to \pi^+ \pi^0 \pi^0 \eta$, the two neutral pions are also indistinguishable and are randomly assigned labels $\pi_1^0$ and $\pi_2^0$; the corresponding set of ten invariant-mass-squared variables—$M_{\pi^+\pi_1^0}$, $M_{\pi^+\pi_2^0}$, $M_{\pi_1^0\pi_2^0}$, $M_{\pi_1^0\pi_2^0\pi^+}$, $M_{\pi^+\eta}$, $M_{\pi_1^0\eta}$, $M_{\pi_2^0\eta}$, $M_{\pi_1^0\pi^+\eta}$, $M_{\pi_2^0\pi^+\eta}$, and $M_{\pi_1^0\pi_2^0\eta}$—is used as input.

Given balanced training samples drawn from $f_{\mathrm{bkg}}(p) \propto B_{\epsilon}(p)\, \epsilon(p)\, R_4(p)$ and $f_{\mathrm{PHSP}}(p) \propto \epsilon(p)\, R_4(p)$, the classifier’s output probabilities satisfy $P_{\mathrm{bkg}}(p)/P_{\mathrm{PHSP}}(p) \propto B_{\epsilon}(p)$~\citep{Cranmer:2015bka}.
We therefore define the background weight as $w(p) = P_{\mathrm{bkg}}(p)/P_{\mathrm{PHSP}}(p)$, which directly encodes the true background dynamics relative to phase space. In the amplitude fit, where the background PDF is proportional to $B_{\epsilon}(p)\, \epsilon(p)\, R_4(p)$ and the common factor $\epsilon(p) R_4(p)$ is absorbed into the normalization (as in Eq.~(1) of the main text), we simply use $f_B(p) \propto w(p)$ as the background model for data events in the signal region.


\section{III. Symmetry Considerations in Amplitude Construction}

\textbf{Note that:} (1) 
The amplitude for the decay $D^{+} \to \pi^+ \pi^{+(0)} \pi^{-(0)} \eta$ must be symmetrized under the exchange of the two identical $\pi^{+(0)}$ mesons to satisfy Bose-Einstein statistics.
For the intermediate resonance processes involving $a_0(980)^{+(0)} \to \eta \pi^{+(0)}$, we construct two amplitudes corresponding to the two possible assignments of the observed $\pi^{+(0)}$ momenta ($\pi^{+(0)}_1$, $\pi^{+(0)}_2$) to the decay products of the resonance, and then coherently superpose these amplitudes with a relative $+$ sign to respect Bose symmetry.

(2) For the $D^+\to \eta(1405)/f_1(1285)/f_1(1420)\pi^+$ decays with the subsequent decays $\eta(1405)/f_1(1285)/f_1(1420) \to a_0(980)^\pm \pi^\mp$, the isospin symmetry relation requires the decay amplitudes for charge-conjugate final states ($a_0(980)^+\pi^-$ versus $a_0(980)^-\pi^+$) to have identical magnitudes and phases, as constrained by the Clebsch-Gordan coefficients in Table~\ref{tab:CG coefficient}.

\begin{table}[htbp]
    \caption{
    Isospin relations and amplitude combinations for $D^+\to X\pi^+$ with $X \to a_0(980)^\pm\pi^\mp$, where $X = \eta(1405)$, $f_1(1285)$, or $f_1(1420)$, and their amplitudes are decomposed into charge-specific amplitudes ($A_1$, $A_2$) and their isospin-symmetric combination $A = A_1 + A_2$.}
    \label{tab:CG coefficient}
    \begin{center}
    \begin{tabular} {l c c}
        \hline \hline
        Index & Amplitude & Relation \\
        \hline
        $A_1$ & $D^+\to \eta(1405)\pi^+, \eta(1405)(\to a_0(980)^+ \pi^-)$ & \\
        $A_2$ & $D^+\to \eta(1405)\pi^+, \eta(1405)(\to a_0(980)^- \pi^+)$ & \\
        $A$  & $D^+\to \eta(1405)\pi^+, \eta(1405)(\to a_0(980)^\pm \pi^\mp)$ & $A_1 + A_2$\\
        \hline
        $A_1$ & $D^+\to f_1(1285)\pi^+, f_1(1285)(\to a_0(980)^+ \pi^-)$ & \\
        $A_2$ & $D^+\to f_1(1285)\pi^+, f_1(1285)(\to a_0(980)^- \pi^+)$ & \\
        $A$  & $D^+\to f_1(1285)\pi^+, f_1(1285)(\to a_0(980)^\pm \pi^\mp)$ & $A_1 + A_2$\\
        \hline
        $A_1$ & $D^+\to f_1(1420)\pi^+, f_1(1420)(\to a_0(980)^+ \pi^-)$ & \\
        $A_2$ & $D^+\to \eta(1405)\pi^+, f_1(1420)(\to f_1(1420) \pi^+)$ & \\
        $A$  & $D^+\to f_1(1420)\pi^+, f_1(1420)(\to a_0(980)^\pm \pi^\mp)$ & $A_1 + A_2$\\
        \hline \hline
    \end{tabular}
    \end{center}
\end{table}

\section{IV. Systematic uncertainties of amplitude analysis}

The systematic uncertainties on the phase ($\phi$) and fit fraction (FF) for each amplitude in the amplitude analysis for $D^+\to\pi^+\pi^+\pi^-\eta$ and $D^+\to\pi^+\pi^0\pi^0\eta$ are summarized in Tables~\ref{tab:pwa_total_syserr for DpToEta3Pi} and \ref{tab:pwa_total_syserr for DpToEta2Pi0Pi}, respectively. The total uncertainty of $\phi$ or FF for each amplitude is obtained by the quadratic sum of individual contributions.

\begin{table*}[htbp]
  \caption{The relative systematic uncertainties (in units of the corresponding statistical uncertainties) on $\phi$ and FF for each amplitude in $D^{+} \to \pi^+ \pi^+ \pi^- \eta$. 
  Sources include: (I) amplitude model, (II) effective barrier radii of resonances, (III) fit bias, (IV) experiment effect, and (V) background.}
  \label{tab:pwa_total_syserr for DpToEta3Pi}
  \centering
  \begin{tabular}{lccccccc}
    \hline \hline
    &\multicolumn{7}{c}{Source} \\
    \hline
    Amplitude  &\ &I  &II  &III  &IV  &V &Total\\
    \hline
$D^+ \to a_0(980)^+ \rho(770)^0$ & FF & 0.91 & 0.05 & 0.07 & 0.00 & 0.12 & 0.92 \\  \hline\multirow{2}{*}{$D^{+}\to a_0(980)^{+}f_0(500)$} & $\phi$ & 2.12 & 0.29 & 0.13 & 0.00 & 0.32 & 2.17 \\ & FF & 0.90 & 0.01 & 0.04 & 0.02 & 0.32 &0.96 \\ \hline\multirow{2}{*}{$D^+\to \eta(1405)(\to a_0(980)^+\pi^-)\pi^+$} & $\phi$ & 1.10 & 0.32 & 0.20 & 0.01 & 0.24 & 1.19 \\ & FF & 0.94 & 0.30 & 0.02 & 0.02 & 0.16 &1.00 \\ \hline\multirow{2}{*}{$D^+\to \eta(1405)(\to a_0(980)^-\pi^+)\pi^+$} & $\phi$ & 1.10 & 0.32 & 0.20 & 0.00 & 0.24 & 1.19 \\ & FF & 0.98 & 0.31 & 0.02 & 0.01 & 0.16 &1.04 \\ \hline\multirow{2}{*}{$D^+\to f_1(1285)(\to a_0(980)^+\pi^-)\pi^+$} & $\phi$ & 2.03 & 0.07 & 0.09 & 0.01 & 0.39 & 2.07 \\ & FF & 0.63 & 0.23 & 0.10 & 0.00 & 0.22 &0.71 \\ \hline\multirow{2}{*}{$D^+\to f_1(1285)(\to a_0(980)^-\pi^+)\pi^+$} & $\phi$ & 2.03 & 0.07 & 0.09 & 0.01 & 0.39 & 2.07 \\ & FF & 0.77 & 0.33 & 0.10 & 0.01 & 0.23 &0.87 \\ \hline\multirow{2}{*}{$D^+\to f_1(1420)(\to a_0(980)^+\pi^-)\pi^+$} & $\phi$ & 1.95 & 0.31 & 0.19 & 0.01 & 0.30 & 2.01 \\ & FF & 1.39 & 0.28 & 0.05 & 0.01 & 0.08 &1.42 \\ \hline\multirow{2}{*}{$D^+\to f_1(1420)(\to a_0(980)^-\pi^+)\pi^+$} & $\phi$ & 1.95 & 0.31 & 0.19 & 0.00 & 0.30 & 2.01 \\ & FF & 1.47 & 0.27 & 0.05 & 0.04 & 0.08 &1.50 \\ \hline\multirow{2}{*}{$D^+\to [a_0(980)^+\pi^-]_S\pi^+$} & $\phi$ & 2.76 & 0.30 & 0.04 & 0.00 & 0.37 & 2.80 \\ & FF & 1.12 & 0.16 & 0.10 & 0.00 & 0.27 &1.17 \\ \hline\multirow{2}{*}{$D^+\to [a_0(980)^-\pi^+]_S\pi^+$} & $\phi$ & 2.76 & 0.30 & 0.04 & 0.00 & 0.37 & 2.80 \\ & FF & 1.06 & 0.16 & 0.10 & 0.00 & 0.27 &1.11 \\ 
\hline \hline
    \end{tabular}
\end{table*}

\begin{table*}[htbp]
  \caption{The relative systematic uncertainties (in units of the corresponding statistical uncertainties) on $\phi$ and FF for each amplitude in $D^{+} \to \pi^+ \pi^0 \pi^0 \eta$. 
  Sources include: (I) amplitude model, (II) effective barrier radii of resonances, (III) fit bias, (IV) experiment effect, and (V) background.}
  \label{tab:pwa_total_syserr for DpToEta2Pi0Pi}
  \centering
  \begin{tabular}{lccccccc}
    \hline \hline
    &\multicolumn{7}{c}{Source} \\
    \hline
    Amplitude  &\ &I  &II  &III  &IV  &V &Total\\
    \hline
$D^+ \to a_0(980)^0 \rho(770)^+$ & FF & 0.73 & 0.01 & 0.13 & 0.00 & 0.09 & 0.75 \\  \hline\multirow{2}{*}{$D^+ \to a_0(980)^+ f_0(500)$} & $\phi$ & 0.10 & 0.10 & 0.23 & 0.00 & 0.17 & 0.32 \\ & FF & 0.25 & 0.04 & 0.14 & 0.02 & 0.15 &0.33 \\ \hline\multirow{2}{*}{$D^+ \to \eta(1405)(\to a_0(980)^0\pi^0)\pi^+$} & $\phi$ & 0.76 & 0.06 & 0.18 & 0.01 & 0.07 & 0.79 \\ & FF & 0.33 & 0.21 & 0.08 & 0.02 & 0.05 &0.40 \\ \hline\multirow{2}{*}{$D^+ \to f_1(1285)(\to a_0(980)^0\pi^0)\pi^+$} & $\phi$ & 1.48 & 0.11 & 0.10 & 0.00 & 0.08 & 1.49 \\ & FF & 0.34 & 0.02 & 0.05 & 0.01 & 0.09 &0.36 \\ \hline\multirow{2}{*}{$D^+ \to f_1(1420)(\to a_0(980)^0\pi^0)\pi^+$} & $\phi$ & 1.70 & 0.14 & 0.19 & 0.01 & 0.15 & 1.72 \\ & FF & 0.48 & 0.18 & 0.08 & 0.00 & 0.07 &0.53 \\ \hline\multirow{2}{*}{$D^+ \to [a_0(980)^0\pi^0]_S\pi^+$} & $\phi$ & 1.41 & 0.15 & 0.20 & 0.01 & 0.09 & 1.43 \\ & FF & 0.32 & 0.08 & 0.05 & 0.01 & 0.09 &0.34 \\ 
\hline \hline
    \end{tabular}
\end{table*}

\section{V. Tag-side Information for Branching Fraction Measurements}

The relevant information used in the BF measurements is summarized in Tables~\ref{tab:signal efficiency for DpToEta3Pi} and \ref{tab:signal efficiency for DpToEta2Pi0Pi}, which list the $\Delta E$ windows, ST yields in data, ST efficiencies, and DT efficiencies for both $D^+ \to \eta \pi^+ \pi^+ \pi^-$ and $D^+ \to \eta \pi^+ \pi^0 \pi^0$ decay modes.

\begin{table*}[htbp]
    \caption{The energy difference~($\Delta$E) windows, ST yields in data~($N_{\rm{ST}}^{\alpha}$), ST efficiencies~($\epsilon_{\rm{ST}}^{\alpha}$), DT efficiencies~($\epsilon_{\rm{DT}}^{\alpha}$) for each tag mode of $D^+ \to \eta \pi^+ \pi^+ \pi^-$. Note that the efficiencies do not include the BFs of all possible daughter particles.}
    \label{tab:signal efficiency for DpToEta3Pi}
    \begin{center}
    \begin{tabular} {l | c c c c}
        \hline \hline
        Tag mode & $\Delta$E~(MeV) & $N_{\rm{ST}}^{\alpha}~(\times 10^{3})$  & $\epsilon_{\rm{ST}}^{\alpha}(\%)$ & $\epsilon_{\rm{DT}}^{\alpha}(\%)$ \\
        \hline
$D^- \to K^+ \pi^- \pi^-$ & $(-25, 24)$ & $5552.8 \pm 2.5$ & $51.10 \pm 0.01$  & $13.51 \pm 0.01$\\$D^- \to K^+ \pi^- \pi^- \pi^0$ & $(-57, 46)$ & $1723.7 \pm 1.8$  &  $24.40 \pm 0.01$  & $5.61 \pm 0.01$ \\$D^- \to K_S^0 \pi^- $ & $(-25, 26)$ & $656.5 \pm 0.8$ &  $51.42 \pm 0.01$  & $13.77 \pm 0.03$ \\$D^- \to K_S^0 \pi^- \pi^0$ & $(-62, 49)$ & $1442.4 \pm 1.5$ &  $26.45 \pm 0.01$  & $6.42 \pm 0.01$ \\$D^- \to K_S^0 \pi^- \pi^- \pi^+ $ & $(-28, 27)$ & $790.2 \pm 1.1$ &  $29.59 \pm 0.01$ & $7.06 \pm 0.02$  \\
        \hline \hline
    \end{tabular}
    \end{center}
\end{table*}

\begin{table*}[htbp]
    \caption{The energy difference~($\Delta$E) windows, ST yields in data~($N_{\rm{ST}}^{\alpha}$), ST efficiencies~($\epsilon_{\rm{ST}}^{\alpha}$), DT efficiencies~($\epsilon_{\rm{DT}}^{\alpha}$) for each tag mode of $D^+ \to \pi^+ \pi^0 \pi^0 \eta$. Note that the efficiencies do not include the BFs of all possible daughter particles.}
    \label{tab:signal efficiency for DpToEta2Pi0Pi}
    \begin{center}
    \begin{tabular} {l | c c c c}
        \hline \hline
        Tag mode & $\Delta$E~(MeV) & $N_{\rm{ST}}^{\alpha}~(\times 10^{3})$  & $\epsilon_{\rm{ST}}^{\alpha}(\%)$ & $\epsilon_{\rm{DT}}^{\alpha}(\%)$ \\
        \hline
$D^- \to K^+ \pi^- \pi^-$ & $(-25, 24)$ & $5552.8 \pm 2.5$ & $51.10 \pm 0.01$  & $4.08 \pm 0.00$\\$D^- \to K^+ \pi^- \pi^- \pi^0$ & $(-57, 46)$ & $1723.7 \pm 1.8$  &  $24.40 \pm 0.01$  & $1.46 \pm 0.00$ \\$D^- \to K_S^0 \pi^- $ & $(-25, 26)$ & $656.5 \pm 0.8$ &  $51.42 \pm 0.01$  & $4.27 \pm 0.01$ \\$D^- \to K_S^0 \pi^- \pi^0$ & $(-62, 49)$ & $1442.4 \pm 1.5$ &  $26.45 \pm 0.01$  & $1.76 \pm 0.00$ \\$D^- \to K_S^0 \pi^- \pi^- \pi^+ $ & $(-28, 27)$ & $790.2 \pm 1.1$ &  $29.59 \pm 0.01$ & $1.88 \pm 0.01$  \\
        \hline \hline
    \end{tabular}
    \end{center}
\end{table*}

\section{VI. Systematic uncertainties of branching fraction measurements}

The systematic uncertainties in the BF measurements for $D^+\to\pi^+\pi^+\pi^-\eta$ and $D^+\to\pi^+\pi^0\pi^0\eta$ are summarized in Table~\ref{tab:BF systematic}. The total systematic uncertainties are obtained by the quadratic sum of individual contributions.

\begin{table*}[htbp]
    \caption{The relative systematic uncertainties for the BF measurements of  $D^+\to\pi^+\pi^+\pi^-\eta$ and $D^+\to\pi^+\pi^0\pi^0\eta$.}
    \label{tab:BF systematic}
    \begin{center}
    \begin{tabular}{l c c}
        \hline \hline
        $\sigma_{\rm syst.}$(\%) & $D^+ \to \pi^+\pi^+\pi^-\eta$  & $D^+ \to \pi^+\pi^0\pi^0\eta$  \\
        \hline
        $N_{\rm ST}^{\rm tot}$ & 0.3 & 0.3 \\
        $\pi^{\pm}$ tracking & 0.3 & 0.1 \\
        $\pi^{\pm}$ PID & 0.1 & 0.1 \\
        $\pi^0/\eta$ reconstruction & 0.3 & 0.8 \\
        $\Delta E_{\rm sig}$ requirement & 0.5 & 0.4 \\
        Amplitude model & 0.3 & 0.7 \\
        2D fit & 0.3 & 0.8 \\
        Cited BFs & 0.5 & 0.5 \\
        MC statistics & 0.1 & 0.1 \\
        \hline
        Total & 1.0 & 1.5 \\
        \hline \hline
    \end{tabular}
    \end{center}
\end{table*}

%
\bibliography{References.bib}

\end{document}